# Single molecule capturer by doped monatomic carbon chain

Zheng-Zhe Lin[1)†] and Xi Chen[2)]

1) *School of Science, Xidian University, Xi'an 710071, P. R. China*

2) *Department of Applied Physics, School of Science, Xi'an Jiaotong University, Xi'an 710049, P.R. China*

[†] Corresponding Author. E-mail address: linzhengzhe@hotmail.com

**Abstract** - The B-doped monatomic carbon chain has fine molecular capture ability to $H_2O$ and especially to $NO_2$, better than other doped monatomic carbon chains. At 300 K and 1 atm, the capture probability of the B-doped monatomic carbon chain is appreciable even in a $NO_2$ concentration of 1 p.p.m., and the adsorbates' influence on the quantum transport is notable for the detection. In contrast, the pure monatomic carbon chain shows its invulnerability to $N_2$, $O_2$, $H_2O$, $NO_2$, CO and $CO_2$, and it is incapable for molecule capturing due to too low adsorption ability and weak response on the quantum conductance. In the investigation of these issues, a statistic mechanical model [*EPL* **94**, 40002 (2011); *Chin. Phys. Lett.* **29**, 2012 (080501)] was extended to predict the adsorption and desorption rates of molecules on nanodevices. The theoretical foundation of this model was further discussed and its accuracy was verified by molecular dynamics simulations.

## 1. Introduction

The ideal goal of a detection method is to achieve the ultimate sensitivity that an individual quantum can be resolved. In the case of molecular sensors, the quantum is a single atom or molecule. To reach such goal, many efforts were focused on the solid sensors because of their high sensitivity and miniature sizes, making them widely used in many applications. As the next-generation sensors, the detectors made by low-dimensional materials have special high sensitivity because their whole volume is exposed to adsorbates and the effect is maximized. Such detectors were firstly prepared by carbon nanotubes and semiconductor nanowires [1, 2], and their high sensitivity to toxic gases in very low concentration was sought for industrial and environmental monitoring. Then, a new detection method by single-layer graphene was developed [3]. The signal of even an individual molecule can be detected in the strong magnetic field by the Hall Effect of electrons or holes induced by the adsorbed molecules on graphene surface acting as electronic donors or acceptors. For one-dimensional monatomic chains, the influence of adsorbates on the electronic property may be even stronger than that for two-dimensional materials. In recent years, free-standing monatomic carbon chains (MCC) were carved out from single-layer graphene by a high-energy electron beam [4], or unraveled from sharp carbon specimens [5, 6] or carbon nanotubes [7], and the MCCs were predicted to be remarkably stable [8]. Recently, a mechanical procedure was proposed to prepare long pure and doped MCCs [9-11] using as nano rectifier [9] or the medium of tunable infrared laser [12]. It was proposed that the electronic properties of MCCs could be controlled by doping a single atom or a diatomic molecule [9]. As the thinnest natural wires, the MCCs should be also greatly influenced on the electronic properties by the molecules adsorbed on it.

In this work, the adsorption and desorption rates of $N_2$, $O_2$, $H_2O$, $NO_2$, CO and $CO_2$ on pure and doped MCCs were studied by the extension of a newly developed statistic mechanical model [8, 13], and the influence of these adsorbates on the current-voltage curves were investigated by quantum transport calculations. The pure MCC was found incapable to be a molecular capturer due to too low adsorption ability and the weak response on the quantum conductance. In contrast, the B-doped MCC has fine capture ability to $H_2O$ and remarkably to $NO_2$. In the ambient of 300 K and 1 atm, the molecular capture probability is appreciable and the electric signal is notable even in a $NO_2$ concentration of 1 p.p.m..

**2. Theoretical model and methodology**

Generally, elementary processes correspond to some individual events that one or two "key atoms" cross over a static barrier $E_0$. For the adsorption of molecules on nanodevices, a basic event takes place when a molecule hits the nanodevice with a translational kinetic energy $\varepsilon \geq E_0$ and a specific orientation for the key atom in the molecule and the nanodevice to contact with each other. At room temperature or above, the thermal De Broglie wavelength of the molecule is at least 100 times smaller than its size, and then the molecular motions can be understood in classical pictures. By the classical ensemble theory, the distribution of molecular translational kinetic energy is Boltzmann, i.e. $f(\varepsilon) \sim \varepsilon^{1/2} e^{-\varepsilon/k_B T}$. On this basis, a simple model was built for the prediction of bimolecular chemical reaction rates [13]. Here, a similar idea was provided for predicting the molecular adsorption rate on nanodevices. By the Boltzmann distribution, the probability for the molecular translational kinetic energy $\varepsilon \geq E_0$ reads

$$P = \frac{1}{Z} \int_{E_0}^{+\infty} \varepsilon^{1/2} e^{-\varepsilon/k_B T} d\varepsilon, \tag{1}$$

where $Z = \int_0^{+\infty} \varepsilon^{1/2} e^{-\varepsilon/k_B T} d\varepsilon = \sqrt{\pi}(k_B T)^{3/2}/2$ the partition function. So, for the molecules at a concentration $c$, the corresponding adsorption rate should be

$$\Gamma = \frac{\sigma v c}{2Z} \int_{E_0}^{+\infty} \varepsilon^{1/2} e^{-\varepsilon/k_B T} d\varepsilon, \tag{2}$$

where $\sigma$ the effective cross-section of the nanodevice and $v = \sqrt{2k_B T/\pi M}$ the average thermal velocity of molecule along the cross-section normal. The factor 1/2 is because only half of the molecules move towards the cross-section. It should be noted that $\sigma$ is not equal to the geometry cross-section $S$ of the key atoms in the nanodevice [Fig. 1(a)], but instead $\sigma = S\Omega_m/4\pi$, where $\Omega_m$ the solid angle opened by the key atom in the molecule. In practical applications, $\Omega_m$ can be estimated by the atomic distance to the molecular mass center and atomic covalent radius [13].

For the elementary processes within a system, e.g. the molecular desorption from a nanodevice, a basic event takes place when the key atom in a potential valley crosses over the barrier with a kinetic energy $\varepsilon$ at the valley bottom larger than $E_0$. The atomic kinetic energy distribution is determined by $f(\varepsilon) = \sum_i f_i(\varepsilon) e^{-E_i/k_BT} / \sum_i e^{-E_i/k_BT}$, here $f_i(\varepsilon)$ is the kinetic energy distribution of the quantum state $E_i$. As an example, $f(\varepsilon)$ of an atom in a Cl$_2$ molecule is shown in Fig. 1(b). At room temperature or above, the distribution $f(\varepsilon)$ turns into the classical one. For a classical system including $N$ atoms, the total energy $E = \vec{p}_1^2/2m_1 + ... + \vec{p}_N^2/2m_N + V(\vec{x}_1,...,\vec{x}_N)$ and the kinetic energy distribution of the i$^{th}$ atom reads

$$\begin{aligned}f(\varepsilon) &= \int \delta[\vec{p}_i^2/2m_i - \varepsilon] e^{-E/k_BT} d\vec{p}_1...d\vec{p}_N d\vec{x}_1...d\vec{x}_N / \int e^{-E/k_BT} d\vec{p}_1...d\vec{p}_N d\vec{x}_1...d\vec{x}_N \\ &= \int \delta[\vec{p}_i^2/2m_i - \varepsilon] e^{-\vec{p}_i^2/2m_ik_BT} d\vec{p}_i / \int e^{-\vec{p}_i^2/2m_ik_BT} d\vec{p}_i \\ &= \varepsilon^{1/2} e^{-\varepsilon/k_BT}/Z\end{aligned} \quad (3)$$

This Boltzmann distribution for atomic kinetic energy is in very good agreement with MD simulations [8]. In most cases, the atomic kinetic energy ($\sim k_BT$) at the valley bottom is significantly smaller than $E_0$, and the atom vibrates many times within the valley before crossing over the barrier. With a vibration frequency $\Gamma_0$, the rate for the atomic event reads

$$\Gamma = \frac{\Gamma_0}{Z} \int_{E_0}^{+\infty} \varepsilon^{1/2} e^{-\varepsilon/k_BT} d\varepsilon. \quad (4)$$

For a given $\varepsilon$, the oscillation period $\tau(\varepsilon) = \sqrt{m} \int d\vec{x}/2[\varepsilon - V(\vec{x})]$ along the minimum energy path (MEP) can be determined by $V(\vec{x}) = \int \vec{F}(\vec{x}) \cdot d\vec{x}$, where $\vec{F}(\vec{x})$ is the force felt by the key atom at position $\vec{x}$ [8]. With the corresponding oscillation frequency $v(\varepsilon)=1/\tau(\varepsilon)$, the averaged frequency reads

$$\Gamma_0 = \frac{\int_0^{E_0} v(\varepsilon) \varepsilon^{1/2} e^{-\varepsilon/k_BT} d\varepsilon}{\int_0^{E_0} \varepsilon^{1/2} e^{-\varepsilon/k_BT} d\varepsilon}. \quad (5)$$

For two-key-atoms processes, an event takes place when the atomic kinetic energy sum $\varepsilon_1+\varepsilon_2 \geq E_0$, and the corresponding rate should be

$$\Gamma = \frac{\Gamma_0}{Z^2} \iint_{\varepsilon_1+\varepsilon_2 \geq E_0} \varepsilon_1^{1/2} \varepsilon_2^{1/2} e^{-(\varepsilon_1+\varepsilon_2)/k_BT} d\varepsilon_1 d\varepsilon_2$$
$$= \frac{\Gamma_0}{2(k_BT)^3} \int_{E_0}^{+\infty} \varepsilon^2 e^{-\varepsilon/k_BT} d\varepsilon \quad . \tag{6}$$

In our previous work [8], Eq. (4) and (6) were verified by MD simulations and successfully applied to predict the stability of MCC-graphene junction and bonds in the MCC body, reproducing results in good agreement with the experimental data.

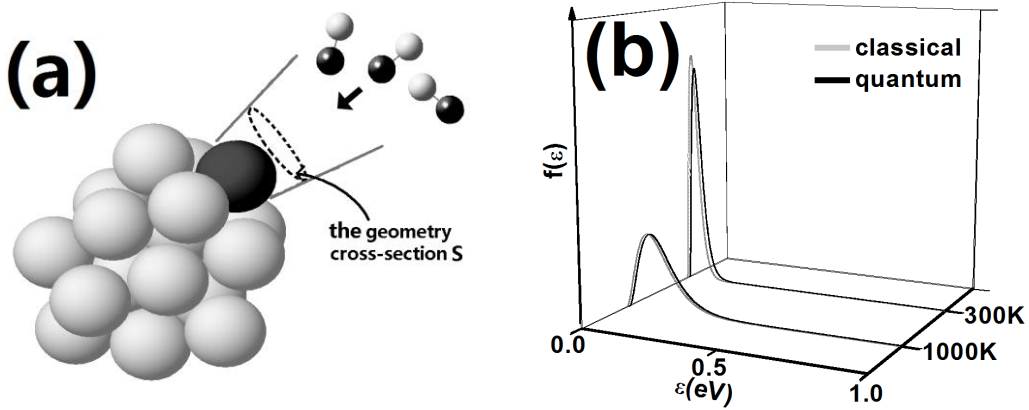

Fig. 1 (a) The geometry cross-section $S$ of the key atom in the nanodevice; (b) The kinetic energy distribution $f(\varepsilon)$ of an atom in a $Cl_2$ molecule at 300 and 1000 K, by classical (gray lines) and quantum mechanics (black lines), respectively.

To investigate the molecular adsorption and desorption rates on MCCs, *ab initio* calculations were performed for a 10-atom pure or doped MCC bridged between two graphene lattices in the armchair [Fig. 2(a)] or zigzag [Fig. 2(b)] orientation with the blacked atoms fixed. Simulations were carried out for the pure, B-doped, N-doped and BN-doped MCC coupling with $N_2$, $O_2$, $H_2O$, $NO_2$, $CO$ and $CO_2$ molecule. The geometry optimizations, reaction barriers $E_0$, MEPs and the forces $F(\bar{x})$ felt by the key atom were calculated using the pseudo reaction coordinate method on level of density functional theory (DFT) via the Gaussian 03 package [14] with the 6-31G(d,p) basic set and the newly developed hybrid X3LYP functional including dispersion interactions [15], which is considered more accurate than other functionals in the potential surface and MEP calculations. Canonical modes for the geometries of potential minima and transition states were calculated to confirm the optimization results.

For quantum transport, the simulation system was setup by appending semi-infinite graphene electrodes in the armchair [Fig. 2(a)] or zigzag [Fig. 2(b)] orientation to the two terminals of the pure or doped MCC, with or without the adsorbed molecule. Calculations were performed by the TRANSIESTA package [16], using non-equilibrium Green's function [17] at the level of Perdew-Burke-Ernzerhof parameterized generalized-gradient approximation [18] with Troullier-Martins pseudopotential [19] and localized double-zeta polarized basis set in order to preserve a correct description of π-conjugated bonds. The mesh cutoff energy is 150 Ry, and a $k$-point sampling of $1\times8\times100$ was used. For a bias voltage $V_b$ applied on the electrodes, the current is given by Landauer-Büttiker equation [16]

$$I = \frac{2e}{h}\int T(E,V_b)[f_L(E-E_F-\frac{eV_b}{2})-f_R(E-E_F+\frac{eV_b}{2})]dE, \qquad (7)$$

where $T(E,V_b)$ the scattering coefficient of the band state at energy $E$, $E_F$ the Fermi energy of the electrodes and $f_L$ and $f_R$ the Fermi-Dirac distribution functions of both electrodes, respectively. As buffer layers, enough graphene lattices in the scattering region are essential to screen the induced electric field between two electrodes. Our calculations showed that the transport property does not change appreciably when there are at least two buffer layers for the armchair orientation, or three for the zigzag orientation.

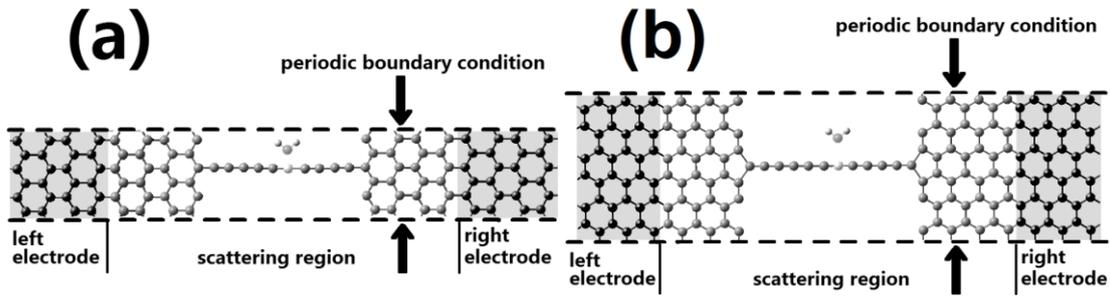

Fig. 2 The simulation system pure or doped MCC of a 10 atoms bridged between two graphene lattices in the armchair (a) or zigzag (b) orientation, with the blacked atoms fixed.

**3. Results and discussion**

**3.1 MD simulation**

To verify Eq. (2), molecular dynamics (MD) simulations were performed for molecular adsorption on a monatomic chain. In a periodic cubic box with a side length of 30 Å, the simulation system was set up by putting a 20-atom MCC with its terminals fixed and 33 helium atoms as the buffer gas (BG) along with a diatomic molecule initialized in a random position. The BG atoms were controlled by a thermal bath at temperature $T$. The Brenner potential [20, 21] was applied for C-C interactions, and the interaction between the atoms of the diatomic molecule reads

$$V_{mm}(r) = C_1 e^{C_2 r} - C_3 e^{-C_4 r}, \tag{8}$$

with a bond energy of 1.05 eV ($C_1$=9.073×10$^5$ eV, $C_2$=10.925 Å$^{-1}$, $C_3$=3.514 eV and $C_4$=0.764 Å$^{-1}$). A modified Leonard-Jones potential

$$V_{cm}(r) = D_1/r^{12} - D_2/r^6 + D_3/r^3 \tag{9}$$

was designed for the interaction between C and the atoms of the diatomic molecule ($D_1$=3.028×10$^3$ eV·Å$^{12}$, $D_2$=3.177×10$^2$ eV·Å$^6$ and $D_3$=33.348 eV·Å$^3$), providing an adsorption barrier $E_{0a}$=1.052 eV. These parameters for the artificially constructed potential Eq. (8) and (9), were adjusted to let the adsorption and desorption happen within the time scale of MD simulations. Because our model does not depend on the specific form of interaction potential, it is suitable for the adsorption and desorption progress of any diatomic molecule on a monatomic chain, and the chosen parameters for Eq. (8) and (9) do not affect the verification of the model. For $T$=700~2000 K, the average time for once event is in the range of several μs to several hundreds of ps, and simulations were performed repeatedly at every temperature until the average adsorption rate changed below 5% (about 10 times for low temperature and 10$^3$ of times for high temperature). The geometry cross-section $\sigma$ of the 20-atom MCC is estimated as $S = 20 \times 2\pi dL = 1101.07$ Å$^2$, where $d$= 6.74 Å the distance from the molecular mass center to the MCC axis where the barrier $E_0$ appears, and $L$=1.30 Å the C-C bond length in the MCC [Fig. 3(a)]. For a diatomic molecule, the solid angle taken by an atom in the molecule gets close to $2\pi$, and for the two atoms $\sigma = S\Omega_m/4\pi \approx S$. By the results, the adsorption rates predicted by Eq. (2) are in good agreement with MD [Fig. 3(b)]. It is worth noting that the model is also applicable to triatomic or polyatomic molecules because Eq. (2) is independent of the molecular geometry.

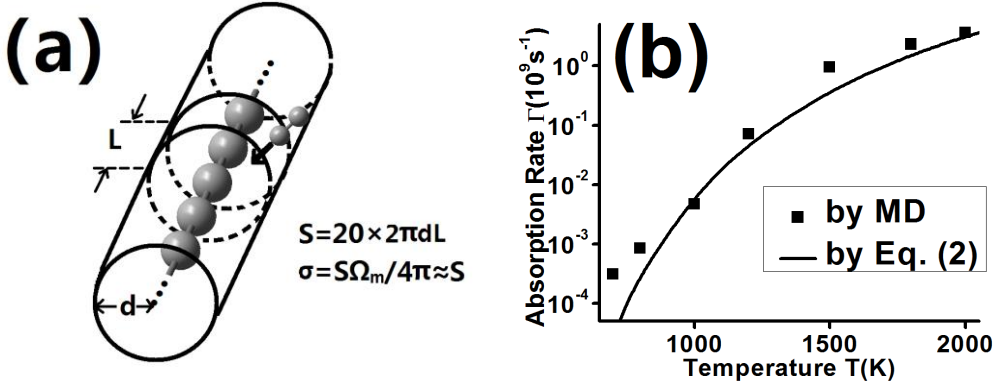

Fig. 3 The cross-section $\sigma$ for the adsorption of molecule (a) described by the artificially constructed potential Eq. (8) and (9), and the corresponding adsorption rate (b) on a 20-atom MCC.

### 3.2 DFT calculation for pure MCC

Along the MEPs of $NO_2$, CO and $CO_2$ molecule approaching the pure MCC, the potential drops to a valley of 0.017~0.028 eV without barriers, while only repulsive interactions were found for $N_2$ and $H_2O$ [Table 1]. The molecular coverage of the MCC is estimated as follows. For the barrierless adsorptions, Eq. (2) becomes $\Gamma_a = \sigma vc/2$. The desorption event takes place when the bond between the MCC and the molecule breaks, i.e. the kinetic energy sum $\varepsilon_1+\varepsilon_2$ of the two bonding atoms is larger than $E_0$, and Eq. (6) should be applied. The molecule can go away from the MCC along the radial direction and two tangential directions perpendicular to the chain axis, and indeed, three equivalent paths were found in the MEP calculations. In the $NO_2$ ambient of 300 K and 1 atm, by $\sigma \approx 28.0$ Å$^2$ the molecular adsorption rate on one C atom is about $\Gamma_a = 6.4 \times 10^8$ s$^{-1}$. By $E_{0d}$=0.024 eV and $\Gamma_0 = 1.0 \times 10^{12}$ s$^{-1}$, the corresponding desorption rate was estimated to be $\Gamma_d = 2.8 \times 10^{12}$ s$^{-1}$, and so, in equilibrium the ratio of C atoms covered by $NO_2$ molecules is $R = \Gamma_a/(\Gamma_a + \Gamma_d) \approx 0.02\%$. At 1000 K, the ratio even decreases to $R \approx 0.01\%$. Such weak interaction presents the invulnerability of MCCs to these molecules, and they have little influence on the quantum transport. The $I$-$V_b$ curves are shown in Fig. 4(a) and (b), for the graphene electrodes in the armchair and zigzag orientation, respectively. Similar situations are also found for CO and $CO_2$. Such little molecular coverage probability and little difference of the $I$-$V_b$ curve between a solo MCC and a MCC with an adsorbed molecule means that the pure MCC could not be used as a capturer or detector for these molecules.

|  | $NO_2$ | CO | $CO_2$ | $N_2$ | $H_2O$ | $O_2$ |
|---|---|---|---|---|---|---|
| adsorption barrier $E_{0a}$ (eV) | 0 | 0 | 0 | - | - | 0.93 |
| desorption barrier $E_{0d}$ (eV) | 0.024 | 0.017 | 0.028 | - | - | 0.38 |

Table 1. The adsorption and desorption barrier of some molecules against a pure MCC, in which "-" means unable to be adsorbed.

For $O_2$ molecule, a barrier $E_{0a}$=0.93 eV for the adsorption and $E_{0d}$=0.38 eV for the desorption [Table 1] were found in the MEP calculation [Fig. 4(c)], respectively, corresponding to an adsorption and desorption rate much slower than $NO_2$, CO and $CO_2$ due to these barriers. In the ambient of 1 atm $O_2$ at 300 K, it takes $\tau_a = 1/\Gamma_a$=222 h and $\tau_d = 1/\Gamma_d$=346 h for a C atom capturing and releasing one $O_2$ molecule, respectively [Fig. 4(d)]. Note that although $E_{0a}$>$E_{0d}$, the adsorption time $\tau_a$ is shorter than the desorption time $\tau_d$ because of the high $O_2$ concentration $c$. In equilibrium the coverage ratio of $O_2$ on the MCC is up to $R = \Gamma_a/(\Gamma_a + \Gamma_d) \approx 60.9\%$. But such status cannot be achieved in a short time due to too long $\tau_a$ and $\tau_d$. At higher temperature, $\tau_a$ becomes much shorter than $\tau_d$ in orders of magnitude, and so, the coverage of $O_2$ sharply increases. For example, at 400 K it takes $\tau_a$=132 s for a C atom capturing an $O_2$, but $\tau_d$=15 h for the releasing progress [Fig. 4(d)]. In this case, the MCC will be rapidly oxidized because the coverage of $O_2$ becomes $R = \Gamma_a / (\Gamma_a + \Gamma_d) \approx 99.8\%$. So, the pure MCC is not suitable for $O_2$ capturing either at room temperature or higher temperature.

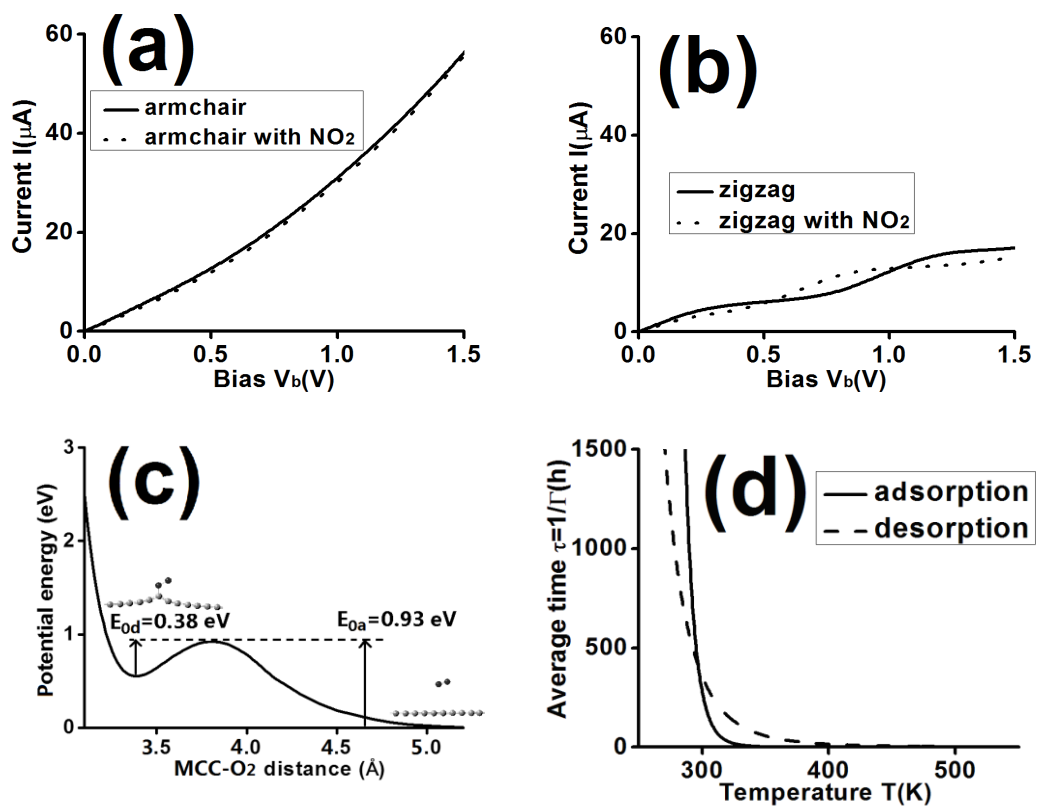

Fig. 4 The $I$-$V_b$ curve for a solo MCC and a MCC with an adsorbed $NO_2$ molecule, with the graphene electrodes in the armchair (a) or zigzag (b) orientation; The potential energy profile along the MEP with the changing distance from the $O_2$ mass center to the MCC axis (c); The average time $\tau = 1/\Gamma$ taken for a C atom in the MCC capturing (solid line) and releasing (dashed line) a molecule in the ambient of 1atm $O_2$ (d).

**3.3 DFT calculation for B- and N-doped MCC**

The molecular adsorption ability could be enhanced by doping one B atom in the MCC because of the bonding by the injecting electrons from the molecule to the empty orbital of the B atom. According to the result, such bonding was found for $H_2O$ and $NO_2$ molecule. For $H_2O$, the unshared pair of electrons on the O atom couples with the B atom, providing an adsorption barrier $E_{0a}$=0.75 eV and a desorption barrier $E_{0d}$=1.12 eV [Table 2]. For a longer MCC, $E_{0a}$ becomes a little lower. At the saturated vapor pressure of $H_2O$ at 300 K (3.6 kPa, corresponding to a concentration of $1.4 \times 10^{-3}$ mol/L), it takes $\tau_a = 1/\Gamma_a$=3.7 h for the B atom to capture a $H_2O$ molecule, and $\tau_d = 1/\Gamma_d = 12$ minutes for the H$_2$O molecule staying on the B atom, which is long enough for detecting the captured molecule [Fig. 5(a)]. Although $E_{0a} < E_{0d}$, $\tau_a$ is longer than $\tau_d$ due to the low H$_2$O concentration $c$. At higher temperature, although the time $\tau_a$ for the capturing is much shorter, the detection becomes unfeasible because $\tau_d$ decreases by orders of magnitude and the captured molecule cannot stay for an appreciable time. For example, at 400 K it takes $\tau_a = 13$ s for once capturing, but the H$_2$O molecule only stays for $\tau_d = 26$ ms on the B atom. When the temperature is below 250 K, $\tau_a$ becomes much larger than that of 300 K in orders of magnitude, indicating that the molecule capturing event could hardly happen at low temperature. Therefore, the B-doped MCC could be a H$_2$O capturer only at a temperature near 300 K. For quantum transport, the adsorption of H$_2$O leads to an obvious reduction in the current $I$ for the graphene electrodes no matter in the armchair [Fig. 5(c)] or zigzag [Fig. 5(d)] orientation, which acts as a notable detectable signal of the B-doped MCC capturing a H$_2$O molecule.

|  | NO$_2$ | CO | CO$_2$ | N$_2$ | H$_2$O |
|---|---|---|---|---|---|
| adsorption barrier $E_{0a}$ (eV) | 0 | 0 | 0 | 0 | 0.75 |
| desorption barrier $E_{0d}$ (eV) | 1.27 | 0.001 | 0.012 | 0.021 | 1.12 |

Table 2. The adsorption and desorption barrier of some molecules against a B-doped MCC.

The adsorption of NO$_2$, CO, CO$_2$ and N$_2$ molecule on the B atom of the doped MCC was found barrierless [Table 2]. For CO, CO$_2$ and N$_2$, the adsorbed molecule leaves the MCC quickly with a high rate $\Gamma_d$ because of too small $E_{0d}$. For NO$_2$, the corresponding adsorption rate $\Gamma_a = \sigma vc/2$ is much faster than that of H$_2$O in orders of magnitude. In the NO$_2$ ambient of 1 atm and the temperature range of $T=300$~1000 K, the time $\tau_a = 1/\Gamma_a = 2 \times 10^{-3}$~$3 \times 10^{-3}$ μs means that the adsorption of NO$_2$ on the B atom takes place at soon. The desorption barrier, i.e. the adsorption energy, was found to be $E_{0d} = 1.27$ eV. At 300 K, the time $\tau_d = 1/\Gamma_d = 20$ h for the captured NO$_2$ molecule staying on the B atom means that the molecule can hardly escape [Fig. 5(b)]. $\tau_d$ decreases with increasing temperature [Fig. 5(b)], and the desorption event, i.e. $\tau_d \ll \tau_a$, happens at temperatures above 1000 K. At 300 K and 1 atm, even when the $NO_2$ concentration decreases to 1 p.p.m. (i.e. a partial pressure of 0.1 Pa) the time $\tau_a$=1.9 ms for capturing a $NO_2$ molecule is still short. So, the B-doped MCC should be a very good capturer for $NO_2$ even in very low concentration. But notable reduction in the quantum transport current $I$ was only found for the graphene electrodes in the zigzag orientation [Fig. 5(d)], while no obvious change in the $I$-$V_b$ curve was found for the armchair one [Fig. 5(c)].

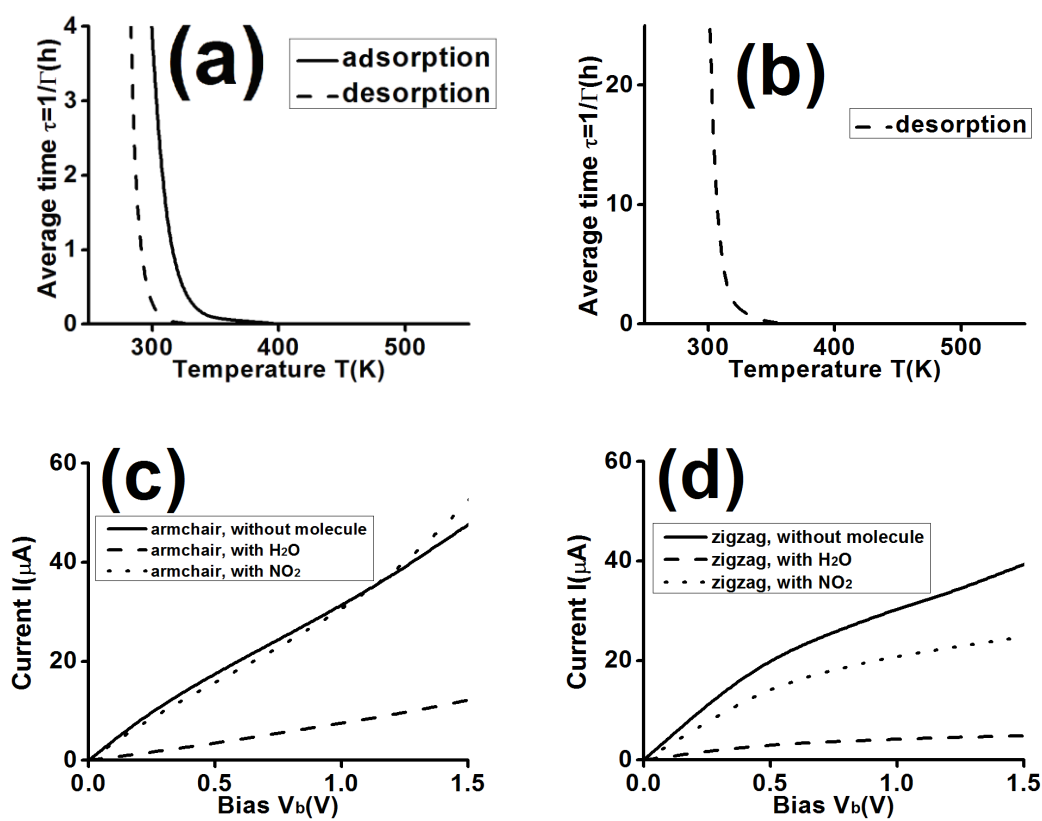

Fig. 5 The average time $\tau$=1/$\Gamma$ taken for the B atom in the doped MCC capturing a molecule in the $H_2O$ vapor of 3.6 kPa (solid line), or releasing an adsorbed $H_2O$ molecule (dashed line) (a); $\tau$ for the B atom releasing an adsorbed $NO_2$ molecule (b); The $I$-$V_b$ curve for a solo B-doped MCC or with an adsorbed $H_2O$ or $NO_2$ molecule on the B atom, with the graphene electrodes in the armchair (c) or zigzag (d) orientation.

The MCC doped by a N atom or a BN diatomic molecule has much lower molecule capture ability than the B-doped MCC. The desorption barrier $E_{0d}$ is in the range of 0.1~0.4 eV for the captured $H_2O$ and $NO_2$ molecule leaving away from the N-doped or BN-doped MCC, while repulsive interactions were found for $N_2$, CO and $CO_2$. Such low $E_{0d}$ corresponds to a very small $\tau_d$ in the order of magnitude of $10^{-7}$~$10^{-3}$ μs at room temperature, indicating that the N-doped and BN-doped MCC could not be used as molecule capturers.

**3.4 Conditions for good molecular capturer**

A good molecular should have short molecular adsorption time $\tau_a=1/\Gamma_a$ and long desorption time $\tau_d=1/\Gamma_d$. For general molecules, the molecular mass and collision cross-section to one atom in the monatomic chain could be roughly estimated as $M\approx10$~50 amu and $\sigma\approx10$~50 Å$^2$, respectively, and by Eq. (5) $\Gamma_0$ generally has a value in the range of $10^{13}$~$10^{14}$ s$^{-1}$. So, $\tau_a$ and $\tau_d$ can be estimated by Eq. (2) and (6). At 300 K, in the 1 atm ambient of the molecule $\tau_a$ is less than several hours if $E_{0a}<0.8$ eV, and $\tau_d$ is longer than dozens of minutes if $E_{0d}>1.1$ eV. For B-doped MCC, $H_2O$ molecule against the B atom just fits this condition, and $NO_2$ molecule could be well captured by the B atom because $E_{0a}=0$ and a $E_{0d}$ larger than that of $H_2O$. In contrast, pure or N-doped MCC could not be good molecular capturer due to too low $E_{0d}$.

**4. Summary**

In this work, a statistic mechanical model [8, 13] was extended to the predict the adsorption and desorption rates of small molecules on nanodevices. This model is based on the fact that the kinetic energy distribution of atoms or molecules always obeys $\varepsilon^{1/2}e^{-\varepsilon/k_BT}$, and its theoretical foundation was further investigated. The accuracy of corresponding extension was verified by MD simulations. By *ab initio* calculations, the extended model was applied on predicting the adsorption and desorption rates of $N_2$, $O_2$, $H_2O$, $NO_2$, CO and $CO_2$ on the pure and doped MCC. By the result, the pure MCC is incapable to be a molecular capturer due to low adsorption probability to $N_2$, $H_2O$, $NO_2$, CO and $CO_2$ molecules at room temperature, and the influence of these molecules on the current-voltage curves is weak due to the poor adsorption. For $O_2$, at room temperature the corresponding adsorption rate is too slow, while at higher temperature the MCC will be rapidly oxidized. In contrast, the B-doped MCC has fine capture ability to $H_2O$ and remarkably to $NO_2$, better than the N-doped or BN-doped MCC, and the reduction of the quantum transport current by the captured molecule is notable, which could be a detectable electronic signal of the molecule capturing. At room temperature and 1 atm, even in a $NO_2$ concentration of 1 p.p.m. the change in the current-voltage curves of the B-doped MCC is appreciable and the captured $NO_2$ could stay for a long time on the B atom, indicating that the B-doped MCC could be a very good single molecule capturer and detector for $NO_2$.

**Acknowledgements**

This work was supported by the Fundamental Research Funds for the Central Universities.